\documentclass[10pt,twocolumn,twoside]{IEEEtran}

\usepackage{cite}
\usepackage{amsmath,amssymb,amsfonts}
\usepackage{amsthm}
\usepackage{graphicx}
\usepackage{array}
\usepackage{url}
\usepackage{xcolor}
\usepackage{multirow}
\usepackage{booktabs}
\usepackage{algorithm}
\usepackage{algpseudocode}
\usepackage[caption=false,font=normalsize,labelfont=sf,textfont=sf]{subfig}

\def\BibTeX{{\rm B\kern-.05em{\sc i\kern-.025em b}\kern-.08em
    T\kern-.1667em\lower.7ex\hbox{E}\kern-.125emX}}
\newtheorem{theorem}{Theorem}
\newtheorem{lemma}{Lemma}
\newtheorem{definition}{Definition}
\newtheorem{assumption}{Assumption}
\newtheorem{remark}{Remark}

\newcommand{\R}{\mathbb{R}}
\newcommand{\N}{\mathcal{N}}
\newcommand{\T}{\mathcal{T}}
\newcommand{\bp}{\boldsymbol{p}}
\newcommand{\bu}{\boldsymbol{u}}
\newcommand{\bx}{\boldsymbol{x}}
\newcommand{\bz}{\boldsymbol{z}}
\newcommand{\bw}{\boldsymbol{w}}
\newcommand{\bmu}{\boldsymbol{\mu}}
\newcommand{\balpha}{\boldsymbol{\alpha}}
\newcommand{\bgamma}{\boldsymbol{\gamma}}
\newcommand{\blambda}{\boldsymbol{\lambda}}
\newcommand{\bxi}{\boldsymbol{\xi}}
\newcommand{\bpsi}{\boldsymbol{\psi}}
\newcommand{\brho}{\boldsymbol{\rho}}

\newcommand{\bc}{\boldsymbol{c}}
\newcommand{\bg}{\boldsymbol{g}}
\newcommand{\bh}{\boldsymbol{h}}

\newcounter{defcounter}
\newenvironment{pequationP}{%
\addtocounter{equation}{-1}
\refstepcounter{defcounter}
\renewcommand\theequation{{\bf{P}}}
\begin{equation}}
{\end{equation}}

\newenvironment{Lpequationi}{%
\addtocounter{equation}{-1}
\refstepcounter{defcounter}
\renewcommand\theequation{{\bf{LP$_i$}}}
\begin{equation}}
{\end{equation}}

\begin{document}

\title{Social Welfare and Equilibrium in Distributed Microgrid Prosumer Coordination Under a Mixed Competitive-Cooperative Setting}

\author{Mahan Fakourifard, \IEEEmembership{Member, IEEE,} Elizabeth L. Ratnam, \IEEEmembership{Member, IEEE,}
        and Mingxi Liu, \IEEEmembership{Member, IEEE}%
\thanks{M. Fakourifard and M. Liu are with the Department of Electrical and Computer Engineering, University of Utah, Salt Lake City, UT 84112 USA (e-mail: mahan.fakourifard@utah.edu; mingxi.liu@utah.edu).}
\thanks{E. L. Ratnam is with the Department of Electrical and Computer Systems Engineering, Monash University, Clayton VIC 3168, Australia (e-mail: liz.ratnam@monash.edu)}
}


\maketitle

\begin{abstract}
Price-based distributed prosumer coordination in distribution networks is fundamentally challenging when the aggregator pursues grid-level goals that couple prosumers' decisions, because the Lagrangian fails to separate and nodal prices are undefined under conventional dual decomposition. This paper develops a novel method that embeds mixed competitive-cooperative objectives exactly into prices, which are split into energy, congestion, and grid-objective components, closing the gap between the social welfare optimum and the competitive equilibrium. To compute the equilibrium, a distributed optimization algorithm is developed based on the projected Arrow-Hurwicz-Uzawa scheme. Explicit step-size rules are developed for the algorithm to guarantee $\mathcal{O}(1/\sqrt{L})$ and $\mathcal{O}(1/L)$ ergodic convergence rates. Simulations on an IEEE 33-bus feeder that is connected to prosumers and electric vehicle fleets validate the accuracy and efficacy of the proposed methods.
\end{abstract}


\section{Introduction}
\IEEEPARstart{C}{ontemporary} distribution feeders are evolving into active networks in which consumers who own distributed energy resources (DERs), including photovoltaic (PV) arrays, battery storage, and electric vehicles (EVs), both consume and supply energy \cite{sandelic2022reliability}. To engage these prosumers in grid management, market-based coordination mechanisms are increasingly used to align individual interests with system-level goals \cite{zhou2017incentive}, facilitating voltage regulation \cite{AmritPaudel2021}, frequency support \cite{King2021Solving}, and demand-side flexibility \cite{tao2021data}. The rapid growth of behind-the-meter PV and EVs is turning passive feeders into bilateral exchanges in which thousands of small assets must be coordinated in near real time. Centralized control of such assets is neither scalable nor privacy-preserving, motivating the concept of transactive energy, where prices steer self-interested prosumers toward system-level goals \cite{hu2017transactive}. The crux lies in the design of the price scheme where selfish responses provably realize a desired global outcome. Price-based coordination does not by itself guarantee privacy, since many transactive designs still require prosumers to report bids, cost functions, or device parameters \cite{kok2016society,chen2017demand}. Privacy therefore requires a price-based scheme in which prosumers share only their tentative injection schedules and keep their payoffs, local generation and demand, and device states private.

A well-established price-based scheme for local energy markets is based on competitive equilibrium (CE) \cite{Li2020} rooted in microeconomics \cite{mas1995microeconomic}. A CE is a set of prices at which each self-interested agent maximizes its own payoff and the market clears, guaranteeing Pareto efficiency under convexity. When payoffs are quasilinear, Pareto efficiency is equivalent to maximizing total social welfare. Selfish responses to CE prices therefore realize the welfare-maximizing outcome. Moreover, the Arrow-Debreu construction casts a CE as the outcome of a game with a price-setting coordinator \cite{arrow1954existence}, so a CE can be reached through a price-mediated negotiation where agents exchange prices and quantities rather than private data \cite{herves2020market}. However, the open question is whether such prices exist and whether they can be computed when prosumers face network operating constraints, dynamic device states, and grid-level objectives that couple them all.

Distributed grid asset coordination has been pursued along four main lines, each of which leaves part of the present problem unaddressed. Hierarchical EV charging frameworks \cite{khaki2018hierarchical} split the task into network- and vehicle-level layers but treat the coordinating price as an exogenous input, so the resulting schedules are not certified to maximize social welfare. Consensus-based schemes \cite{nimalsiri2021coordinated,rahbari2014cooperative,sorin2019consensus} reach agreement without a central coordinator, but they typically do not model distribution voltage limits, so their schedules may violate these limits on heavily loaded radial feeders. Duality-based market mechanisms \cite{Li2020,salehi2021social} make the price-optimality link explicit for static or single-snapshot loads. The CE was extended in \cite{salehi2024competitive} to dynamic loads, such as batteries whose state of charge (SoC) evolves, under voltage constraints. However, these mechanisms maximize a social welfare that is separable across prosumers and do not handle a grid-level objective that couples all of them. Distribution locational marginal pricing (DLMP) \cite{papavasiliou2018dlmp} delivers locational, congestion-aware prices, but existing theories presume a separable objective and do not extend to grid-level goals that couple all prosumers. The defining challenge is that such a grid-level goal, for instance, load tracking at the point of common coupling (PCC), is inseparable: It couples all prosumers' injections, so dual decomposition \cite{boydvandenberghe2004,palomar2006tutorial} yields no per-prosumer price. Hence, this paper is positioned to answer the open question and fill the aforementioned gaps.

In this paper, we develop a novel DLMP scheme that coordinates prosumers toward a mixed competitive-cooperative grid-level objective. In the proposed scheme, each nodal price carries the substation energy cost, the network voltage limits, and the marginal value of the grid-level objective, while the device dynamics of each prosumer stay in its local problem. Compared with the aforementioned existing pricing schemes, the proposed method attains a CE, and hence the maximum social welfare, under voltage limits, dynamic device states, and an inseparable grid-level objective simultaneously, while keeping all private data local. The contribution of this paper is three-fold: 
\begin{itemize}
\item We, for the first time, develop a DLMP scheme that inherently embeds an inseparable grid-level objective into the nodal prices and guarantees the social-welfare optimum and CE. The objective enters every price through a single scalar per time step without compromising optimality and with linear communication overhead. 

\item The scheme prices finite-horizon device dynamics jointly with network voltage limits and the inseparable grid-level objective, differentiating prosumers in both time and location. 

\item A distributed algorithm that only requires the exchange of prices and injection schedules and imposes linear communication and computational complexities is developed to compute the CE. Explicit step-size rules are developed to guarantee $\mathcal{O}(1/\sqrt{L})$ and $\mathcal{O}(1/L)$ ergodic convergence rates after $L$ iterations.
\end{itemize}

The remainder of the paper is organized as follows. Section \ref{sec:formulation} formulates the problem. Section \ref{sec:duality} establishes the zero duality gap and the price equivalence. Section \ref{sec:algorithm} develops the projected Arrow-Hurwicz-Uzawa (AHU) scheme and its convergence guarantee. Section \ref{sec:results} reports the case study. Section \ref{sec:conclusion} concludes this paper. 

\section{Problem Formulation}
\label{sec:formulation}
Consider a radial distribution feeder whose substation is labeled as node $0$ and treated as the slack bus. Each of the other nodes hosts one prosumer, indexed by $i\in\N=\{1,\dots,n\}$, so $i$ identifies a prosumer and its node interchangeably. A node hosting multiple prosumers is handled by treating their aggregate as a single prosumer. Each prosumer has one controllable device, i.e., an EV or an EV fleet aggregated into a single equivalent battery. Each prosumer's rooftop PV array and inflexible load are not controlled, and together with the EV, they sit behind the meter. Operation spans a horizon of $T$ intervals of length $\Delta$, indexed by $\T=\{0,\dots,T-1\}$, and all power quantities indexed by $t$ are averaged over interval $t$. We denote by $p_i(t)$ the net power that prosumer $i$ injects into the grid, with $p_i(t)>0$ indicating export and $p_i(t)<0$ indicating import. The net power is collected in $\bp(t)=[p_1(t),\ldots,p_n(t)]^{\mathsf{T}}\in\R^{n}$. We denote by $p_0(t)$ the power that the feeder imports from the transmission grid, with $p_0(t)>0$ indicating power flows into the feeder.

\subsection{System and Network Model}
\label{subsec:prosumer}
Suppose prosumer $i$ is connected to an EV with SoC $x_i(t)\in\R$, evolving as
\begin{equation}\label{eq:dyn}
x_i(t+1)=x_i(t)+\eta_i\,\Delta\,u_i(t),
\end{equation}
where $x_i(0)=x_i^0$ is the initial SoC, $u_i(t)\in\R$ denotes the charging power, with a positive value, or the discharging power, with a negative value, and $\eta_i$ is the charging efficiency. Each EV imposes the operating limits as
\begin{subequations}\label{eq:devbox}
\begin{align}
\underline{x}_i\le x_i(t)\le\overline{x}_i&,\ t\in\T\cup\{T\},\label{eq:socbox}\\
-\overline{u}_i\le u_i(t)\le\overline{u}_i&,\ t\in\T .\label{eq:ubox}
\end{align}
\end{subequations}
To protect the battery, the SoC bounds are restricted to an interior window $0<\underline{x}_i<\overline{x}_i<C$, where $C$ is the battery capacity. Let $a_i(t)\in\R$ be the prosumer's net surplus power, i.e., local generation minus inflexible load. In each interval, the power that prosumer $i$ injects into the grid plus the power it uses to charge its EV cannot exceed its surplus, indicating
\begin{subequations}\label{eq:surplus}
\begin{align}
p_i(t)+u_i(t)&\le a_i(t), \label{eq:surplus_a}\\
-\overline{p}_i\le p_i(t)&\le\overline{p}_i,\label{eq:pbox}
\end{align}
\end{subequations}
where $\overline{p}_i>0$ is the connection limit of prosumer $i$. Power balance at the meter gives equality in \eqref{eq:surplus_a} when all PV output is used, and the inequality permits PV curtailment.
Prosumer $i$ values its charging schedule through two terms: a \emph{running payoff} $f_i(u_i(t))$ scores the charging decision in each interval and a \emph{terminal payoff} $\Phi_i(x_i(T))$ scores the energy actually delivered by the end of the horizon. Specifically, we adopt the standard quadratic forms of \cite{salehi2024competitive}
\begin{subequations}\label{eq:payoffs}
\begin{align}
f_i(u_i(t))&=-c_u\,u_i(t)^2,\\
\Phi_i(x_i(T))&=-c_T\left(x_i(T)-x_i^{\mathrm{ref}}\right)^2,
\end{align}
\end{subequations}
where $c_u,c_T>0$ weigh charging effort and terminal-SoC deviation, respectively, and $x_i^{\mathrm{ref}}$ is the departure target SoC.

With the LinDistFlow approximation \cite{baran1989distflow,farivar2013equilibrium}, the squared voltage magnitude deviation at node $i$ is linear in the nodal injections, represented as
\begin{equation}\label{eq:lindist}
v_i(t)-v_0=\sum_{k\in\N}R_{ik}\,p_k(t)+\sum_{k\in\N}X_{ik}\,q_k(t),
\end{equation}
where $v_i(t)$ is the squared voltage magnitude at node $i$, $v_0$ its value at the substation, $q_k(t)$ the reactive injection at node $k$, and
\begin{equation}\label{eq:Rdef}
R_{ik}=2\!\!\sum_{(h,l)\in\mathcal{P}_i\cap\mathcal{P}_k}\!\!r_{hl},\qquad
X_{ik}=2\!\!\sum_{(h,l)\in\mathcal{P}_i\cap\mathcal{P}_k}\!\!x_{hl},
\end{equation}
in which $r_{hl}$ and $x_{hl}$ are the resistance and reactance of the line joining nodes $h$ and $l$, and $\mathcal{P}_i$ denotes the set of lines on the path from the substation to node $i$. We make the following assumption on the feeder's reactive power.
\begin{assumption}\label{ass:q}
The reactive injections $q_k(t)$, $\forall k\in\N$, are exogenous and known to the aggregator. \hfill $\square$
\end{assumption}

Assumption \ref{ass:q} is standard in DER scheduling: Residential inverters are operated at a fixed power factor, so $q_k(t)$ is determined by the local load rather than chosen by the prosumer. Hence, the reactive term of \eqref{eq:lindist} is a known constant, leaving a constraint that is linear in the active injections alone. The admissible voltage band then becomes
\begin{equation}\label{eq:volt}
\underline{v}_i\le\sum_{k\in\N}R_{ik}\,p_k(t)\le\overline{v}_i,\qquad \forall i\in\N .
\end{equation}
The power balance over the feeder then reads
\begin{subequations}\label{eq:balance}
\begin{align}
p_0(t)+\sum_{i\in\N}p_i(t)&=0,\label{eq:bal}\\
\underline{p}_0\le p_0(t)&\le\overline{p}_0 .\label{eq:p0box}
\end{align}
\end{subequations}
The substation incurs a convex supply cost
\begin{equation}\label{eq:feedercost}
C_0\left(p_0(t)\right)=\pi_0(t)\,p_0(t)+c_0\,p_0(t)^2 ,
\end{equation}
where $\pi_0(t)$ is the time-of-use import price and $c_0>0$ weighs a quadratic term that models substation losses and ramping. 

\subsection{Social-Welfare Problem with an Inseparable Objective}
\label{subsec:grid}
The aggregator pursues an objective that couples all prosumers through their aggregate injection $\sum_i p_i(t)$. The leading example is PCC load tracking, i.e., $p_0(t)$ is required to follow a reference $P_b(t)$ chosen by the aggregator, captured by
\begin{equation}\label{eq:g}
g\left(\bp(t)\right)=-\mu_g\left(P_b(t)-\sum_{i\in\N}p_i(t)\right)^{2},
\end{equation}
with weight $\mu_g>0$. Crucially, $g(\cdot)$ depends on the prosumers only through their linear combination and
\begin{equation}\label{eq:gderiv}
\frac{\partial g}{\partial p_i}\left(\boldsymbol{p}(t)\right)=2\mu_g\left(P_b(t)-\sum_{j\in\N}p_j(t)\right)
\end{equation}
is identical for all prosumers.
In the rest of this paper, we drop the time index to stack a quantity over a time horizon and drop the node index to stack over the prosumers. For example, $\bu_i=[u_i(0),\dots,u_i(T-1)]^{\mathsf{T}}\in\R^{T}$ is the charging trajectory of prosumer $i$, $\bp(t)\in\R^{n}$ collects the injections at time $t$, and $\bu=[\bu_1^{\mathsf{T}},\dots,\bu_n^{\mathsf{T}}]^{\mathsf{T}}\in\R^{nT}$ collects all charging trajectories. 
With these notations, the centralized social-welfare problem is defined as
\begin{pequationP}\label{eq:P}
\begin{aligned}
\max_{\,\bu,\bp,\bp_0}\;\; & \sum_{i\in\N}\sum_{t\in\T} f_i\left(u_i(t)\right)+\sum_{i\in\N}\Phi_i\left(x_i(T)\right) \\
&\quad-\sum_{t\in\T}C_0\left(p_0(t)\right)+\sum_{t\in\T}g\left(\bp(t)\right) \\ 
\text{s.t.}\;\; & \eqref{eq:dyn},\ \eqref{eq:devbox},\ \eqref{eq:surplus},\ \eqref{eq:volt},\ \eqref{eq:balance}, \quad \forall i\in\N,\ t\in\T. 
\end{aligned}
\end{pequationP}
Herein, the payoffs $f_i(\cdot)$ and $\Phi_i(\cdot)$ and the local constraints \eqref{eq:dyn}, \eqref{eq:devbox}, and \eqref{eq:surplus} belong to prosumer $i$. The supply cost $C_0(\cdot)$, the grid-level objective $g(\cdot)$, the voltage limits \eqref{eq:volt}, and the power balance \eqref{eq:balance} belong to the aggregator and reach the prosumers only through the nodal prices.

Given a nodal price $\lambda_i(t)$ for the energy injected at node $i$ at time $t$, prosumer $i$ solves the local problem written as
\begin{Lpequationi}\label{eq:LP}
\begin{aligned}
\max_{\bu_i,\bp_i}\;\; & \sum_{t\in\T}\left(f_i\left(u_i(t)\right)+\lambda_i(t)\,p_i(t)\right)+\Phi_i\left(x_i(T)\right) \\
\text{s.t.}\;\; & \eqref{eq:dyn},\ \eqref{eq:devbox},\ \eqref{eq:surplus},\quad \forall t\in\T, 
\end{aligned}
\end{Lpequationi}
which involves only local data and constraints. 

\begin{figure*}[!htb]
\centering
\includegraphics[width=\textwidth]{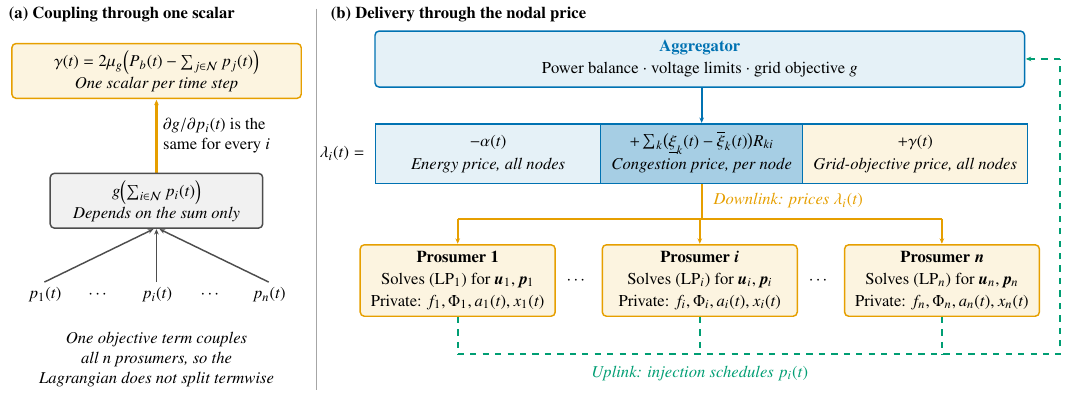}
\caption{Coordination mechanism. (a) The grid-level objective depends on the injections only through their sum, so its partial derivative is the same scalar $\gamma(t)$ for every prosumer. (b) The aggregator broadcasts the nodal price $\lambda_i(t)$; each prosumer solves its local problem using private data; and each prosumer returns its injection schedule $p_i(t)$ to the aggregator.}
\label{fig:framework}
\end{figure*}

\subsection{Proposed Framework and Information Structure}
\label{subsec:framework}


This paper proposes a framework that allows the centralized social-welfare problem \eqref{eq:P}
be decomposed into $n$ independent per-prosumer problems \eqref{eq:LP} that are solved locally by individual prosumers, coordinated only by a vector of nodal prices that the aggregator broadcasts. 
The decomposition is challenged by the coupled grid-level objective as it depends on all injections jointly rather than as a separable sum. Moreover, the term $g(\cdot)$ in the Lagrangian of \eqref{eq:P} does not split across prosumers, so standard dual decomposition does not apply directly. In the next section we show that the proposed approach, nevertheless, decomposes the problem exactly, with the entire influence of that coupled objective carried by a single scalar per time step. 

Fig. \ref{fig:framework} summarizes how the proposed framework handles the coupling. Panel (a) shows where the coupling comes from and how it is removed. The grid-level objective $g(\cdot)$ depends on the $n$ nodal injections only through their sum, so its partial derivative with respect to every injection takes the same value $\gamma(t)$. The inseparable term, therefore, enters every local problem through a single scalar per time step. Panel (b) shows how that scalar reaches the prosumers. The aggregator broadcasts a nodal price $\lambda_i(t)$ with three components. The energy and grid-objective components are common to all nodes, while the congestion component varies by node and reflects voltage limits. Each prosumer solves its local problem \eqref{eq:LP} at its own price, using only its private payoffs, surplus, and SoC, and returns its injection schedule $p_i(t)$ to the aggregator, which uses these schedules to update the prices. Only prices and injection schedules are exchanged, so each prosumer's payoffs, surplus, and SoC remain private.

\section{Duality and Competitive Equilibrium}
\label{sec:duality}
\subsection{Nodal Price via Duality}
\label{subsec:price}
Collect all decision variables in $\bz={[\bu^{\mathsf{T}},\bp^{\mathsf{T}},\bp_0^{\mathsf{T}}]}^{\mathsf{T}}\in\R^{(2n+1)T}$ and denote by $W(\bz)$ the objective function of \eqref{eq:P}. Define the per-prosumer local constraint set as
\begin{equation}\label{eq:Xi}
\mathcal{X}_i=\left\{(\bu_i,\bp_i):\ \eqref{eq:dyn},\ \eqref{eq:devbox},\ \eqref{eq:surplus}\ \text{hold}\right\},
\end{equation}
which is convex and compact and involves only the data of prosumer $i$. 
We introduce the following standard assumption.  

\begin{assumption}\label{ass:reg}
\textup{(A1)} For each $i$, $f_i(\cdot)$ and $\Phi_i(\cdot)$ are concave and continuously differentiable, $C_0(\cdot)$ is convex and continuously differentiable, and $g(\cdot)$ in \eqref{eq:g} is concave and continuously differentiable.
\textup{(A2)} The feasible set jointly defined by \eqref{eq:dyn}, \eqref{eq:devbox}, \eqref{eq:surplus}, \eqref{eq:volt}, and \eqref{eq:balance} is nonempty, convex, and compact.
\textup{(A3)} \textup{(Slater)} There exists a point satisfying \eqref{eq:dyn} and \eqref{eq:bal} and all inequality constraints in \eqref{eq:devbox}, \eqref{eq:surplus}, \eqref{eq:volt}, and \eqref{eq:p0box} strictly. \hfill $\square$
\end{assumption}


The partial Lagrangian of \eqref{eq:P} is defined as 
\begin{align}\label{eq:lag}
\mathcal{L}(\bz,\balpha,\Xi)=&\sum_{i,t}f_i(u_i(t))+\sum_i\Phi_i(x_i(T))-\sum_t C_0(p_0(t))\notag\\
&+\sum_t g(\bp(t))-\sum_t\alpha(t)\left(p_0(t)+\sum_i p_i(t)\right)\notag\\
&-\sum_{t,i}\overline{\xi}_i(t)\left(\textstyle\sum_k R_{ik}p_k(t)-\overline{v}_i\right)\notag\\
&-\sum_{t,i}\underline{\xi}_i(t)\left(\underline{v}_i-\textstyle\sum_k R_{ik}p_k(t)\right),
\end{align}
where $\alpha(t)\in\R$ is the multiplier for the balance equation \eqref{eq:bal}, and $\overline{\xi}_i(t)\ge0$ and $\underline{\xi}_i(t)\ge0$ are the multipliers for the upper and lower voltage limits in \eqref{eq:volt} at node $i$, respectively. The voltage multipliers are collected in $\Xi=\{\overline{\xi}_i(t),\underline{\xi}_i(t)\}_{i\in\N,t\in\T}$. The Lagrangian in \eqref{eq:lag} is partial because only the two coupling constraints \eqref{eq:volt} and \eqref{eq:bal} are dualized, while the local sets \eqref{eq:Xi} are retained as local constraints for prosumer $i$ and the substation limit \eqref{eq:p0box} is retained as a local constraint of the aggregator. Retaining these constraints exposes the decomposition, since a Lagrangian in which every constraint were dualized would obscure which terms are private to a prosumer. The only term in \eqref{eq:lag} that is not separable across $i$ is $g(\cdot)$. Its partial derivative \eqref{eq:gderiv} takes the same value for every prosumer, i.e., it is one scalar per time step rather than one number per prosumer per time step, which we denote
\begin{equation}\label{eq:gamma}
\gamma^\star(t)\;:=\;\frac{\partial g}{\partial p_i}\Big|_{\bp^\star(t)}=2\mu_g\left(P_b(t)-\sum_{j\in\N}p_j^\star(t)\right),
\end{equation}
evaluated at the injection profile $\bp^\star(t)$ obtained by solving \eqref{eq:P}. It is identical for all $i$ and computable from the aggregate injection alone, as illustrated in Fig. \ref{fig:framework}. Setting $\partial\mathcal{L}/\partial p_0(t)=0$, which is the stationarity condition for $p_0(t)$ whenever the optimizer lies strictly inside the substation box $[\underline{p}_0,\overline{p}_0]$ of \eqref{eq:p0box}, and using $C_0'(p_0(t))=\pi_0(t)+2c_0p_0(t)$ from \eqref{eq:feedercost} yield the energy price
\begin{equation}\label{eq:energyprice}
\alpha^\star(t)=-C_0'\left(p_0^\star(t)\right)=-\left(\pi_0(t)+2c_0\,p_0^\star(t)\right).
\end{equation}
Herein, $-\alpha^\star(t)$ is exactly the marginal cost of supplying one more unit at the substation. Since $\pi_0(t)>0$, the energy price is positive whenever the feeder imports power, i.e., $p_0^\star(t)\ge0$. Collecting in \eqref{eq:lag} every term that multiplies with $p_i(t)$, including those from $g(\cdot)$, the balance term, and the two voltage terms, gives the coefficient of $p_i(t)$ in prosumer $i$'s share of the Lagrangian, represented as
\begin{equation}\label{eq:price}
\displaystyle\lambda_i^\star(t)=\underbrace{-\alpha^\star(t)}_{\text{Energy}}+\underbrace{\sum_{k\in\N}\left(\underline{\xi}_k^\star(t)-\overline{\xi}_k^\star(t)\right)R_{ki}}_{\text{Congestion}}+\underbrace{\gamma^\star(t)}_{\text{Grid objective}}.
\end{equation}
The price in \eqref{eq:price} is a distribution locational marginal price (DLMP) composed of a system-wide energy component, a locational congestion component, and the marginal value of the inseparable grid objective conveyed by the single scalar $\gamma^\star(t)$. The three components correspond to the three blocks of the price in Fig. \ref{fig:framework}. For the load-tracking objective \eqref{eq:g}, substituting \eqref{eq:gamma} into \eqref{eq:price} gives the grid-objective component $\gamma^\star(t)=2\mu_g(P_b(t)-\sum_j p_j^\star(t))$, which is proportional to the tracking deviation.
This is where the grid-level objective is decoupled: Although $g(\cdot)$ is inseparable, it reaches every local problem only through the single time series $\gamma^\star(t)$, so the aggregator needs the aggregate injection $\sum_j p_j(t)$ and broadcasts one scalar per time step. No prosumer-specific information about $g$ is exchanged, thereby preserving privacy and keeping communication complexity at $\mathcal{O}(nT)$.

\subsection{Social Welfare and Competitive Equilibrium}
We next explicitly define the CE before presenting a theorem about its equivalence with the social welfare solution.
\begin{definition}[Competitive equilibrium]\label{def:ce}
Prices $\{\blambda_i^\star\}$, including an energy price $\balpha^\star$, congestion prices $\Xi^\star=\{\underline{\bxi}^\star,\overline{\bxi}^\star\}\ge0$, and a grid-objective price $\bgamma^\star$, together with an allocation $\bz^\star=(\bu^\star,\bp^\star,\bp_0^\star)$ form a CE if
\begin{enumerate}
\item[(i)] \emph{(prosumers)} for each $i$, $(\bu_i^\star,\bp_i^\star)$ solves \eqref{eq:LP} at price $\blambda_i^\star$ given by \eqref{eq:price};
\item[(ii)] \emph{(import)} $p_0^\star(t)\in\arg\max_{\underline{p}_0\le p_0(t)\le\overline{p}_0}\{-C_0(p_0(t))-\alpha^\star(t)\,p_0(t)\}$ $\forall t\in\T$;
\item[(iii)] \emph{(consistency)} $\gamma^\star(t)=2\mu_g(P_b(t)-\sum_j p_j^\star(t))$ $\forall t\in\T$;
\item[(iv)] \emph{(market clearing)} the allocation is feasible for the two coupling constraints, and the congestion prices are consistent with them: The power balance \eqref{eq:bal} holds exactly at every $t$, the voltage limits \eqref{eq:volt} hold at every node and time, and the complementary-slackness conditions $\overline{\xi}_i^\star(t)\left(\sum_k R_{ik}p_k^\star(t)-\overline{v}_i\right)=0$ and $\underline{\xi}_i^\star(t)\left(\underline{v}_i-\sum_k R_{ik}p_k^\star(t)\right)=0$ hold for all $i\in\N$ and $t\in\T$, so a congestion price is nonzero only where and when the corresponding limit is active. \hfill $\square$
\end{enumerate}
\end{definition}

\begin{theorem}[Equivalence]\label{thm:equiv}
Under Assumption \ref{ass:reg}, $\bz^\star$ solves the social-welfare problem \eqref{eq:P} if and only if there exist $(\balpha^\star,\Xi^\star,\bgamma^\star)$ such that $(\{\blambda_i^\star\},\balpha^\star,\Xi^\star,\bgamma^\star,\bz^\star)$ is a CE in the sense of Definition \ref{def:ce}. In that case, $\bz^\star$ is both the social welfare solution and the CE allocation, and the nodal prices are given by \eqref{eq:price}. \hfill $\blacksquare$
\end{theorem}
\begin{proof}
See \textsc{Appendix} \ref{proof_Th_1}.
\end{proof}

Each of the three terms in \eqref{eq:price} carries a distinct and verifiable meaning at the CE. The energy component $-\alpha^\star(t)$ is a universal marginal cost of substation supply. The congestion component $\sum_k(\underline{\xi}_k^\star(t)-\overline{\xi}_k^\star(t))R_{ki}$ is nonzero only when at least one voltage constraint is active. Its sign is that of $\underline{\xi}_k^\star(t)-\overline{\xi}_k^\star(t)$ weighted by $R_{ki}\ge0$ and therefore records whether an injection at node $i$ relieves or aggravates that limit, and since $R_{ki}$ depends only on the electrical position of node $i$, two prosumers at the same location face the same component and hence the same price. The grid-objective component $\gamma^\star(t)$ is also universal, and it satisfies $\partial\gamma^\star(t)/\partial\left(P_b(t)-\sum_j p_j(t)\right)=2\mu_g>0$, so a larger aggregate deficit raises every prosumer's incentive to inject. Among the three components, only the congestion term is the source of inter-prosumer differentiation. Since it equals prosumer $i$'s marginal contribution to the binding voltage constraints, the prices are cost-reflective, i.e., each prosumer is charged exactly its marginal impact on the shared resources and none cross-subsidizes another, which is the precise sense in which the equilibrium prices are fair.

\section{Distributed Algorithm and Convergence}
\label{sec:algorithm}
This section presents a projected AHU-based distributed primal-dual scheme \cite{arrow1958studies,he2022arrowhurwicz,bin2024ahu} to find the CE. 
The update rules follow the standard projected AHU, and running averages are used for convergence analysis. 
We collect the multipliers in $\Psi=\{\psi_i(t)\}_{i\in\N,t\in\T}$ and $\mathrm{P}=\{\overline{\rho}_i(t),\underline{\rho}_i(t)\}_{i\in\N,t\in\T}$, and stack all multipliers in $\bmu=[\balpha^{\mathsf{T}},\overline{\bxi}^{\mathsf{T}},\underline{\bxi}^{\mathsf{T}},\bpsi^{\mathsf{T}},\overline{\brho}^{\mathsf{T}},\underline{\brho}^{\mathsf{T}}]^{\mathsf{T}}\in\R^{T+5nT}$. Let $\mathcal{M}=\R^{T}\times\R_{\ge0}^{m}$, where $m=5nT$, be the set the multipliers range over. Let $\Pi_{\mathcal{S}}(\cdot)$ denote Euclidean projection onto a closed convex set $\mathcal{S}$.

\subsection{Distributed Algorithm for the Competitive Equilibrium}
With iteration index $\ell$, primal step $\beta_p>0$, and dual step $\beta_d>0$, the projected AHU updates are as follows. Prosumer $i$ (in parallel) ascends its local Lagrangian gradient as
\begin{subequations}\label{eq:primal}
\begin{align}
u_i^{\ell+1}(\tau)&=\Pi_{[-\overline{u}_i,\overline{u}_i]}\Bigl[u_i^\ell(\tau)+\beta_p\bigl(f_i'(u_i^\ell(\tau))\notag\\
&\hspace{-1.2em}+\Phi_i'(x_i^\ell(T))\eta_i\Delta-\psi_i^\ell(\tau)-\eta_i\Delta\!\!\sum_{t>\tau}\!\bigl(\overline{\rho}_i^\ell(t)-\underline{\rho}_i^\ell(t)\bigr)\bigr)\Bigr],\\
p_i^{\ell+1}(\tau)&=\Pi_{[-\overline{p}_i,\overline{p}_i]}\left[p_i^\ell(\tau)+\beta_p\left(\lambda_i^\ell(\tau)-\psi_i^\ell(\tau)\right)\right],
\end{align}
\end{subequations}
where the nodal price and the grid-objective component at iteration $\ell$ are
\begin{subequations}\label{eq:priceiter}
\begin{align}
\lambda_i^\ell(\tau)&=-\alpha^\ell(\tau){+}\sum_{k\in\N}\left(\underline{\xi}_k^\ell(\tau){-}\overline{\xi}_k^\ell(\tau)\right)R_{ki}+\gamma^\ell(\tau),\\
\gamma^\ell(\tau)&=2\mu_g\Bigl(P_b(\tau)-\sum_{j\in\N}p_j^\ell(\tau)\Bigr).
\end{align}
\end{subequations}

The aggregator updates the import at the substation by
\begin{equation}\label{eq:p0update}
p_0^{\ell+1}(\tau)=\Pi_{[\underline{p}_0,\overline{p}_0]}\left[p_0^\ell(\tau)+\beta_p\left(-C_0'(p_0^\ell(\tau))-\alpha^\ell(\tau)\right)\right].
\end{equation}

The aggregator updates the multipliers of the coupling constraints by
\begin{subequations}\label{eq:dual}
\begin{align}
\alpha^{\ell+1}(\tau)&=\alpha^\ell(\tau)+\beta_d\Bigl(p_0^{\ell+1}(\tau)+\sum_{i\in\N}p_i^{\ell+1}(\tau)\Bigr),\\
\overline{\xi}_i^{\ell+1}(\tau)&=\Pi_{\R_{\ge0}}\Bigl[\overline{\xi}_i^\ell(\tau){+}\beta_d\Bigl(\sum_{k\in\N}R_{ik}p_k^{\ell+1}(\tau){-}\overline{v}_i\Bigr)\Bigr],\\
\underline{\xi}_i^{\ell+1}(\tau)&=\Pi_{\R_{\ge0}}\Bigl[\underline{\xi}_i^\ell(\tau){+}\beta_d\Bigl(\underline{v}_i{-}\sum_{k\in\N}R_{ik}p_k^{\ell+1}(\tau)\Bigr)\Bigr].
\end{align}
\end{subequations}

Each prosumer updates its local multipliers by
\begin{subequations}\label{eq:localdual}
\begin{align}
\psi_i^{\ell+1}(\tau)&=\Pi_{\R_{\ge0}}\bigl[\psi_i^\ell(\tau)+\beta_d\bigl(p_i^{\ell+1}(\tau)+u_i^{\ell+1}(\tau)-a_i(\tau)\bigr)\bigr],\\
\overline{\rho}_i^{\ell+1}(t)&=\Pi_{\R_{\ge0}}\left[\overline{\rho}_i^\ell(t)+\beta_d\left(x_i^{\ell+1}(t)-\overline{x}_i\right)\right],\\
\underline{\rho}_i^{\ell+1}(t)&=\Pi_{\R_{\ge0}}\left[\underline{\rho}_i^\ell(t)+\beta_d\left(\underline{x}_i-x_i^{\ell+1}(t)\right)\right],
\end{align}
\end{subequations}
for $\tau\in\T$ and $t\in\{1,\dots,T\}$, where $x_i^{\ell+1}(t)$ follows from $\bu_i^{\ell+1}$ through \eqref{eq:dyn}. The above established algorithm is referred to as \textbf{Algorithm 1}. 


\subsection{Convergence of the Distributed Algorithm}
The convergence analysis concerns the running averages defined as
\begin{equation}\label{eq:avg}
\bar\bz^{\ell}=\frac{1}{\ell}\sum_{s=1}^{\ell}\bz^{s},\qquad \bar\bmu^{\ell}=\frac{1}{\ell}\sum_{s=0}^{\ell-1}\bmu^{s}.
\end{equation}
We first present a lemma on the bounded primal trajectory.
\begin{lemma}[Bounded primal trajectory]\label{lem:bound}
Let
\begin{equation}\label{eq:Zset}
\mathcal{Z}=\left\{\bz\in\R^{(2n+1)T}:\ \eqref{eq:ubox},\ \eqref{eq:pbox},\ \eqref{eq:p0box}\ \text{hold}\ \forall i\in\N,\ t\in\T\right\}
\end{equation}
denote the box constraint set. The primal updates \eqref{eq:primal} and \eqref{eq:p0update} project onto $\mathcal{Z}$. Under Assumption \ref{ass:reg}, every primal iterate $\bz^\ell$ lies in $\mathcal{Z}$, which is convex and compact, and the induced SoC trajectory is bounded. Consequently, for the bounded multiplier set $\mathcal{M}_\rho=\{\bmu\in\mathcal{M}:\|\bmu\|_2\le\rho\}$ with $\mathcal{M}=\R^{T}\times\R_{\ge0}^{m}$, there is a finite $G=G(\rho)$ with $\|\nabla_{\bz}\mathcal{L}(\bz,\bmu)\|_2\le G$ and $\|\nabla_{\bmu}\mathcal{L}(\bz,\bmu)\|_2\le G$ for all $\bz\in\mathcal{Z}$, $\bmu\in\mathcal{M}_\rho$. \hfill $\blacksquare$
\end{lemma}
\begin{proof}
Each primal update in \eqref{eq:primal} and \eqref{eq:p0update} is a Euclidean projection onto a corresponding box constraint. Consequently, $\bz^\ell\in\mathcal{Z}$ for every $\ell$. The set $\mathcal{Z}$ is defined by finitely many lower and upper bounds, hence convex and compact. The affine SoC mapping in \eqref{eq:dyn} sends the bounded $\bu_i$ to a bounded $\bx_i$. The objective is quadratic, and every constraint is affine, thus $\nabla\mathcal{L}$ is continuous and bounded on the compact set $\mathcal{Z}\times\mathcal{M}_\rho$.
\end{proof}

\begin{remark}
Only the box constraints in \eqref{eq:Zset} are enforced at every iteration. The surplus constraint in \eqref{eq:surplus}, the SoC limits \eqref{eq:socbox}, the voltage limits \eqref{eq:volt}, and the power balance \eqref{eq:bal} are dualized, so an individual iterate $\bz^\ell$ violates them in general. Only the infeasibility of the running average $\bar\bz^{\ell}$ vanishes as $\ell$ grows.\hfill $\square$
\end{remark} 

\begin{theorem}[Ergodic convergence and step-size rule]\label{thm:conv}
Let Assumption \ref{ass:reg} hold. Fix an iteration count $L$ and a radius $\rho\ge2\|\bmu^\star\|_2$, and assume that the dual iterates satisfy $\|\bmu^\ell\|_2\le\rho$ for $\ell=0,\dots,L$. Let $G$ be the bound of Lemma \ref{lem:bound} on $\mathcal{Z}\times\mathcal{M}_\rho$, and define
\begin{equation}\label{eq:D}
D^2=\sup_{\bz\in\mathcal{Z}}\|\bz^0-\bz\|_2^2+\sup_{\bmu\in\mathcal{M}_\rho}\|\bmu^0-\bmu\|_2^2 .
\end{equation}
Run Algorithm 1 with the constant steps
\begin{equation}\label{eq:step}
\beta_p=\beta_d=\beta=\frac{D}{2G\sqrt{L}} .
\end{equation}
Then the running averages $\bar\bz^{L}$ and $\bar\bmu^{L}$ satisfy
\begin{equation}\label{eq:gap}
\sup_{\bz\in\mathcal{Z},\;\bmu\in\mathcal{M}_\rho}\left\{\mathcal{L}\left(\bz,\bar\bmu^{L}\right)-\mathcal{L}\left(\bar \bz^{L},\bmu\right)\right\}\;\le\;\frac{2DG}{\sqrt{L}} .
\end{equation}
Consequently, the averaged primal infeasibility and the suboptimality $|W^\star-W(\bar \bz^{L})|$ are both $\mathcal{O}(1/\sqrt{L})$. Moreover, as $L\to\infty$, the limit of every convergent subsequence of $\{(\bar\bz^{L},\bar\bmu^{L})\}$ is a saddle point of $\mathcal{L}$, i.e., an optimal primal-dual pair of \eqref{eq:P}. The induced prices $\blambda_i^\star$ in \eqref{eq:price} support the CE of Theorem \ref{thm:equiv}. \hfill $\blacksquare$
\end{theorem}
\begin{proof}
See \textsc{Appendix} \ref{proof_Th_2}.
\end{proof}

We next present a theorem concerning the extragradient variant of the proposed algorithm. Let $F(\bz,\bmu)=\bigl[-\nabla_{\bz}\mathcal{L}(\bz,\bmu)^{\mathsf{T}},\,\nabla_{\bmu}\mathcal{L}(\bz,\bmu)^{\mathsf{T}}\bigr]^{\mathsf{T}}$ denote the saddle operator of $\mathcal{L}$. Let $\mathcal{A}$ and $\bc_0$ denote the constant matrix and vector such that $\bc(\bz)=\mathcal{A}\bz+\bc_0$, and define $\Lambda=\|\nabla^2_{\bz\bz}\mathcal{L}\|_2+\|\mathcal{A}\|_2$.

\begin{theorem}[Extragradient variant attains $\mathcal{O}(1/L)$]\label{thm:eg}
Replace the updates in each iteration of Algorithm 1 by the extragradient (predictor-corrector) step
\begin{subequations}\label{eq:eg}
\begin{align}
\hat \bz^{\ell}&=\Pi_{\mathcal{Z}}\left[\bz^\ell+\beta\nabla_{\bz}\mathcal{L}(\bz^\ell,\bmu^\ell)\right],\label{eq:eg_hz}\\
\hat\bmu^{\ell}&=\Pi_{\mathcal{M}}\left[\bmu^\ell-\beta\nabla_{\bmu}\mathcal{L}(\bz^\ell,\bmu^\ell)\right],\label{eq:eg_hmu}\\
\bz^{\ell+1}&=\Pi_{\mathcal{Z}}\left[\bz^\ell+\beta\nabla_{\bz}\mathcal{L}(\hat \bz^\ell,\hat\bmu^\ell)\right],\label{eq:eg_z}\\
\bmu^{\ell+1}&=\Pi_{\mathcal{M}}\left[\bmu^\ell-\beta\nabla_{\bmu}\mathcal{L}(\hat \bz^\ell,\hat\bmu^\ell)\right],\label{eq:eg_mu}
\end{align}
\end{subequations}
where \eqref{eq:eg_hz} and \eqref{eq:eg_hmu} form the predictor step and \eqref{eq:eg_z} and \eqref{eq:eg_mu} form the corrector step. Both corrector gradients are evaluated at the predictor point $(\hat\bz^\ell,\hat\bmu^\ell)$. Then, for any step $\beta\le1/(\sqrt{2}\,\Lambda)$, the averages $\bar\bz^{L}={1}/{L}\sum_{\ell=0}^{L-1}\hat\bz^{\ell}$ and $\bar\bmu^{L}={1}/{L}\sum_{\ell=0}^{L-1}\hat\bmu^{\ell}$ of the predictor points satisfy
\begin{equation}\label{eq:eggap}
\sup_{\bz\in\mathcal{Z},\,\bmu\in\mathcal{M}_\rho}\left\{\mathcal{L}\left(\bz,\bar\bmu^{L}\right)-\mathcal{L}\left(\bar \bz^{L},\bmu\right)\right\}\le\frac{D^2}{2\beta L},
\end{equation}
which is $\mathcal{O}(1/L)$ in the iteration count. \hfill $\blacksquare$
\end{theorem}
\begin{proof}
See \textsc{Appendix} \ref{Proof_Th_3}.
\end{proof}

\begin{remark}
 Theorem \ref{thm:eg} improves the rate from $\mathcal{O}(1/\sqrt{L})$ to $\mathcal{O}(1/L)$ at the cost of a second gradient evaluation and a second broadcast per iteration.\hfill $\square$
\end{remark}

\subsection{Implementation of the Distributed Algorithm}
\label{subsec:impl}
We discuss three practical aspects of Algorithm 1.
\begin{enumerate}
\item \emph{Initialization:} The algorithm starts from $\bz^0=0$ and $\bmu^0=0$. In a receding-horizon deployment, the aggregator can instead start from the solution of the previous step, which is expected to be close to the new solution.

\item \emph{Stopping criteria:} Theorem \ref{thm:conv} bounds the saddle-point gap, which cannot be computed because it requires a supremum over $\mathcal{Z}\times\mathcal{M}_\rho$. Instead, the aggregator monitors two quantities of the running average $\bar\bz^{\ell}$. The first is the power-balance residual $r_{\mathrm{bal}}^{\ell}$ defined as
\begin{equation}\label{eq:rbal}
r_{\mathrm{bal}}^{\ell}=\max_{t\in\T}\Bigl|\bar p_0^{\ell}(t)+\sum_{i\in\N}\bar p_i^{\ell}(t)\Bigr|,
\end{equation}
where $\bar p_i^{\ell}(t)$ and $\bar p_0^{\ell}(t)$ are entries of $\bar\bz^{\ell}$. The second is the largest violation of the other dualized constraints $r_{\mathrm{vio}}^{\ell}$ defined as
\begin{align}\label{eq:rvio}
r_{\mathrm{vio}}^{\ell}=&\max_{i\in\N,\,t\in\T}\max\Bigl\{0,\ \bar p_i^{\ell}(t)+\bar u_i^{\ell}(t)-a_i(t),\notag\\
&\bar x_i^{\ell}(t)-\overline{x}_i,\ \underline{x}_i-\bar x_i^{\ell}(t), \textstyle\sum_{k}R_{ik}\bar p_k^{\ell}(t)-\overline{v}_i,\notag\\
&\underline{v}_i-\textstyle\sum_{k}R_{ik}\bar p_k^{\ell}(t)\Bigr\},
\end{align}
where $\bar x_i^{\ell}(t)$ is the SoC that \eqref{eq:dyn} produces from $\bar\bu_i^{\ell}$. Both residuals are scalars because they take the maximum over all nodes and time steps. The algorithm stops when $r_{\mathrm{bal}}^{\ell}\le\epsilon_{\mathrm{bal}}$ and $r_{\mathrm{vio}}^{\ell}\le\epsilon_{\mathrm{vio}}$ for tolerances $\epsilon_{\mathrm{bal}},\epsilon_{\mathrm{vio}}>0$ chosen by the aggregator. Both quantities are bounded by $\|\bc^+(\bar\bz^{\ell})\|_2$ per Theorem \ref{thm:conv}, so after $L$ iterations they are at most $4DG/(\rho\sqrt{L})$ by \eqref{eq:infeas}.


\item \emph{Communication and privacy:} In each iteration, the aggregator sends prices to the prosumers, called the downlink, and each prosumer sends its tentative injection profile to the aggregator, called the uplink. In the downlink, the aggregator broadcasts, for every $t\in\T$, the energy price $\alpha^\ell(t)$, the grid-objective price $\gamma^\ell(t)$, and the $n$ congestion prices, together determining $\lambda_i^\ell(t)$ in \eqref{eq:priceiter}. In the uplink, prosumer $i$ sends $p_i^{\ell+1}(t)$ for all $t\in\T$ that are used by the aggregator to evaluate $\gamma^{\ell+1}(t)$ and the voltage map in \eqref{eq:dual}. Each iteration therefore exchanges $\mathcal{O}(nT)$ scalars in each direction. The prosumers' payoff functions $f_i(\cdot)$ and $\Phi_i(\cdot)$, surplus $a_i(t)$, SoC $x_i(t)$, and device limits are never transmitted. Moreover, the grid-level objective $g(\cdot)$ reaches the prosumers only through $\gamma^\ell(t)$, so its form is not disclosed to them.
\end{enumerate}

\section{Numerical Results}
\label{sec:results}
We validate the proposed framework on a single-phase IEEE 33-bus distribution feeder with respect to the exactness of the equivalence between the social welfare solution and the CE, the zero duality gap, the convergence of the two distributed algorithms, the delivery of the inseparable objective through prices, and the interpretability of the resulting DLMPs.

\subsection{Test System and Data}
\label{subsec:case}
\begin{figure}[!htb]
\centering
\includegraphics[width=0.8\columnwidth]{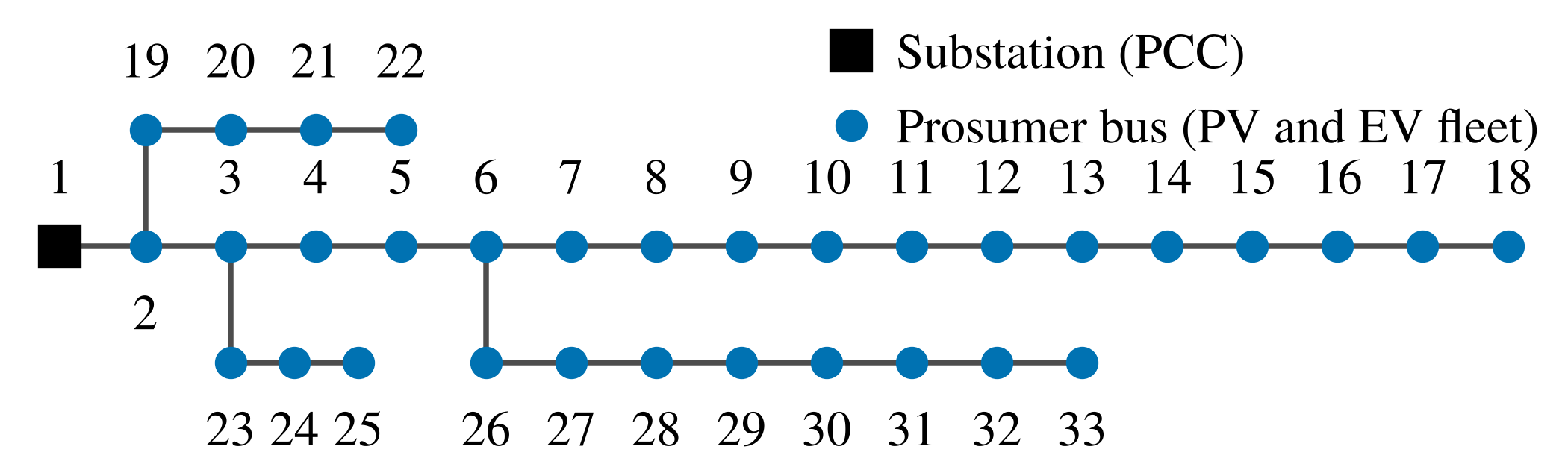}
\vspace*{-3mm}
\caption{The IEEE 33-bus test feeder used in the simulations.}
\label{fig:feeder}
\vspace*{-3mm}
\end{figure}

The $12.66$ kV single-phase distribution feeder \cite{baran1989distflow} is shown in Fig. \ref{fig:feeder}. All quantities are per unit on a $1$ MVA base. The five tie switches are omitted.
One prosumer is connected at each of the $32$ non-substation buses, and each owns a rooftop PV array and an aggregated EV fleet operated as a single equivalent battery. Each vehicle has a $60$ kWh battery, a $7.2$ kW charging limit, and a charging efficiency of $\eta_i=0.95$, and each fleet keeps its SoC within $[0.2C_i,0.85C_i]$ with the target $x_i^{\mathrm{ref}}=0.85C_i$. The fleet size at each bus is proportional to its nominal demand, giving $1,362$ vehicles and $9.81$ MW of charging capacity in total. The payoff weights are $c_u=4\times10^{-3}$ and $c_T=1.0$ per vehicle. The inflexible load at each bus is its nominal demand, scaled to $55\%$ of the peak and shaped by a residential profile. The PV output at each bus follows a clear-sky profile whose peak is twice the peak demand of the bus. A random factor per bus, with a standard deviation of $10\%$ for the load and $12\%$ for the PV, makes the prosumers heterogeneous, and the initial SoC of each fleet is drawn between $25\%$ and $50\%$ of its capacity. The import price $\pi_0(t)$ is a time-of-use tariff between $0.11$ and $0.24$ \$/kWh, and $c_0=10^{-5}$ \$/(kW$^2$h). The horizon is one day at half-hourly resolution, i.e., $T=48$ and $\Delta=0.5$ h. The voltage magnitude at every bus is kept within $[0.95,1.05]$ p.u..

\subsection{Exactness of the Equivalence and the Duality Gap}
\label{subsec:exact}
Theorem \ref{thm:equiv} asserts that the social welfare optimum and the CE coincide exactly. To test this without circularity, the two sides are computed by independent means. First, \eqref{eq:P} is solved directly as a convex quadratic program, giving $W^\star=-9047.2380$ with a primal residual of $1.8\times10^{-9}$. Its multipliers are then used to assemble the price $\blambda^\star$ via \eqref{eq:gamma}-\eqref{eq:price}. Finally, each of the $32$ local problems \eqref{eq:LP} is solved separately at that price. TABLE \ref{tab:compare}
lists the comparisons between the social welfare optimum and the CE, denoted by superscripts $\mathrm{SW}$ and $\mathrm{CE}$, respectively.\begin{table}[!htb]
\centering
\vspace*{-4mm}
\caption{Social Optimum vs.\ Competitive Equilibrium}
\label{tab:compare}
\setlength{\tabcolsep}{3pt}
\begin{tabular}{lr}
\toprule
\textbf{Quantity} & \textbf{Value}\\
\midrule
$\max_{i,t}|u_i^{\mathrm{SW}}(t)-u_i^{\mathrm{CE}}(t)|$ & $1.9\times10^{-6}$ kW\\
$\max_{i,t}|p_i^{\mathrm{SW}}(t)-p_i^{\mathrm{CE}}(t)|$ & $1.9\times10^{-6}$ kW\\
$\max_{i,t}|x_i^{\mathrm{SW}}(t)-x_i^{\mathrm{CE}}(t)|$ & $8.2\times10^{-6}$ kWh\\
Relative payoff gap $\delta_S$ & $1.3\times10^{-7}\%$\\
Relative welfare gap $\delta_W$ & $6.9\times10^{-11}\%$\\
\midrule
$\max_{i,t}|\psi_i^\star(t)/\Delta-\lambda_i^\star(t)|$ & $1.5\times10^{-10}$ \$/kWh\\
Duality gap $q(\bmu^\star)-W(\bz^\star)$ & $1.6\times10^{-7}\$$\\
\bottomrule
\end{tabular}
\vspace*{-1mm}
\end{table} Let $S(\bu)=\sum_{i\in\N}\bigl(\sum_{t\in\T}f_i(u_i(t))+\Phi_i(x_i(T))\bigr)$ denote the aggregate prosumer payoff. The relative gaps in the aggregate prosumer payoff and in the social welfare are defined as
\begin{align}
\delta_S&=\left|\frac{S(\bu^{\mathrm{SW}})-S(\bu^{\mathrm{CE}})}{S(\bu^{\mathrm{SW}})}\right|\times100\%,\label{eq:deltaS}\\
\delta_W&=\left|\frac{W^\star-W(\bz^{\mathrm{CE}})}{W^\star}\right|\times100\%,\label{eq:deltaW}
\end{align}
where $\bz^{\mathrm{CE}}$ is the CE allocation.

From Table \ref{tab:compare}, the two allocations differ by at most $1.9\times10^{-6}$ kW, confirming  Theorem \ref{thm:equiv}. The agreement relies on $c_u>0$, which makes each prosumer's payoff strictly concave. The best response of each prosumer to the equilibrium price is then unique and equals its share of the social optimum. As $c_u$ approaches zero, the best response is not unique. The equivalence still holds, but the local problems may return a different optimal allocation, so the two allocations cannot be compared directly. Table \ref{tab:compare} also confirms two further results. First, at the solution, every injection $p_i(t)$ lies strictly inside its bounds in \eqref{eq:pbox}, so \eqref{eq:kkt_lam} states that the nodal price equals the shadow price of the surplus constraint. In the simulation, the feeder cost, the grid objective, and the payment $\lambda_i(t)p_i(t)$ are multiplied by $\Delta$ to express them in dollars, and the condition becomes $\lambda_i^\star(t)=\psi_i^\star(t)/\Delta$. Over all $1,536$ pairs of prosumer and time step, the price assembled from \eqref{eq:gamma}-\eqref{eq:price} and the multiplier $\psi_i^\star(t)/\Delta$ returned by the solver differ by at most $1.5\times10^{-10}$ \$/kWh. Second, evaluating the dual function at the optimal multipliers gives $q(\bmu^\star){-}W(\bz^\star){=}1.6\times10^{-7}$ \$, which is $1.7\times10^{-9}$ percent of $|W^\star|$. The duality gap is therefore zero up to solver accuracy, as established in Section \ref{sec:duality}.

\subsection{Convergence of the Distributed Schemes}
\label{subsec:conv}
This section presents the convergence analysis of Algorithm 1 and its extragradient variant. Both schemes start from zero primal and dual iterates and run for $4\times10^{5}$ iterations. To offset the differences in the orders of magnitude in the variable updates of  $\bu$, $\bp$, $\bp_0$, and $\bmu$, the primal step for each variable group is $\beta$ multiplied by the largest absolute bound of that group, and the dual step is $\beta$. 
This scaling does not affect the applicability of Theorems \ref{thm:conv} and \ref{thm:eg}. The largest stable step is $\beta=10^{-3}$ for the single-step scheme and $\beta=3\times10^{-4}$ for the extragradient variant, whose step is also bounded by the condition $\beta\le1/(\sqrt2\Lambda)$ of Theorem \ref{thm:eg}.

\begin{figure}[!htb]
\centering
\includegraphics[width=\columnwidth]{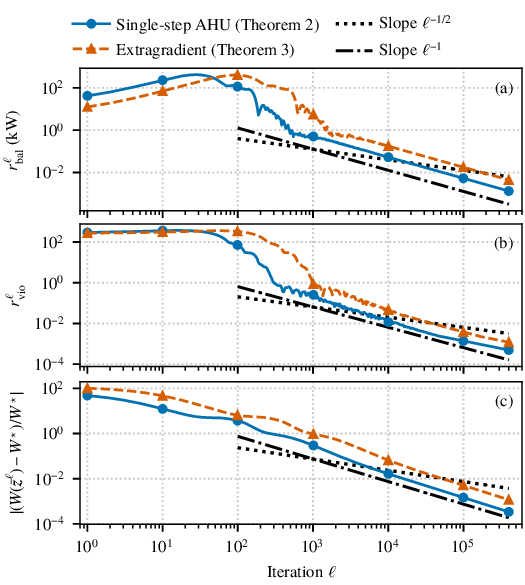}
\vspace*{-8mm}
\caption{Convergence of the running averages of the single-step AHU recursion (Theorem \ref{thm:conv}) and its extragradient variant (Theorem \ref{thm:eg}), measured against the exact solution. (a) Power-balance residual \eqref{eq:rbal}. (b) Largest constraint violation \eqref{eq:rvio}. (c) Relative welfare error.}
\vspace*{-6mm}
\label{fig:conv}
\end{figure}

Fig. \ref{fig:conv} shows three accuracy measures of the running average $\bar\bz^{\ell}$, including residual $r_{\mathrm{bal}}^{\ell}$, violation $r_{\mathrm{vio}}^{\ell}$, and the relative welfare error $|W(\bar\bz^{\ell})-W^\star|/|W^\star|$. For the extragradient variant, the running average is taken over the predictor points, as in Theorem \ref{thm:eg}. After $4\times10^5$ iterations, the single-step scheme attains $r_{\mathrm{bal}}=1.3\times10^{-3}$ kW, $r_{\mathrm{vio}}=5.0\times10^{-4}$ p.u. and a welfare error of $3.4\times10^{-4}$. The extragradient variant attains $4.4\times10^{-3}$ kW, $1.2\times10^{-3}$ p.u. and $1.2\times10^{-3}$, respectively. For both schemes, the largest entry of $r_{\mathrm{vio}}$ comes from the lower voltage limit in \eqref{eq:volt}. On the squared magnitude, the violation of $1.2\times10^{-3}$ p.u. corresponds to less than $7\times10^{-4}$ p.u. in voltage magnitude. The balance residual is negligible compared with the PCC import, which is of the order of $10^3$ kW. Both distributed schemes therefore reach an allocation that satisfies the network constraints to within a negligible margin and whose welfare is within $0.12\%$ of the optimum, without solving \eqref{eq:P} centrally. 

Two remarks are in order. First, the single-step scheme decays faster than its worst-case rate $\mathcal{O}(\ell^{-1/2})$ in all three measures, while the extragradient variant decays at about its rate $\mathcal{O}(\ell^{-1})$. This is expected because those bounds hold for every convex-concave saddle problem with bounded subgradients, whereas the present Lagrangian is quadratic in $\bz$ and affine in $\bmu$, a benign regime. Second, and less expected, the single-step scheme is the better choice for this problem, i.e., it achieves a smaller error in half the wall-clock time because it tolerates a step three times larger and evaluates one gradient per iteration instead of two. The value of Theorem \ref{thm:eg} is therefore the guarantee it provides rather than an empirical speed-up, and the extragradient variant is the safer option when the problem data are less well conditioned than here.

\subsection{Load Tracking, Voltage, and Prices}
\label{subsec:tracking}
\begin{figure}[!htb]
\centering
\vspace*{-3mm}
\includegraphics[width=\columnwidth]{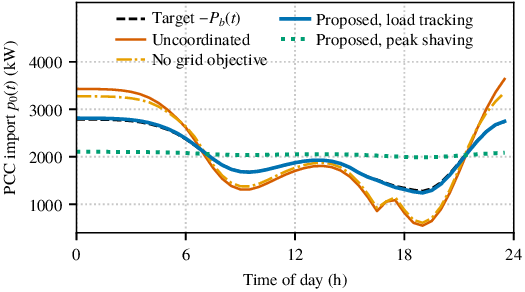}
\vspace*{-7mm}
\caption{Target PCC import and PCC import under different policies.}
\vspace*{-3mm}
\label{fig:tracking}
\end{figure}

\begin{figure}[!htb]
\centering
\vspace*{-1mm}
\includegraphics[width=\columnwidth]{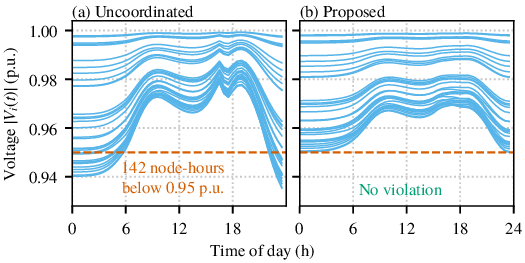}
\vspace*{-7mm}
\caption{Nodal voltage magnitudes over the day under (a) uncoordinated dispatch and (b) the proposed prices.}
\label{fig:voltage}
\end{figure}

The rest of the case study uses PCC load tracking with $\mu_g=5\times10^{-4}$. The PCC import target $-P_b(t)$ is shown in Fig. \ref{fig:tracking}. 
We compare three operating policies on the same data. The first policy is the uncoordinated dispatch, i.e., each prosumer simply minimizes its own cost against the time-of-use tariff. The second policy, termed no grid objective, retains the voltage constraints but sets $\mu_g=0$. The third policy is the proposed approach, i.e., the full problem \eqref{eq:P} with load tracking. Tracking accuracy is measured by the root-mean-square error (RMSE) $\sqrt{{1}/{T}\sum_{t\in\T}(P_b(t)-\sum_{i\in\N}p_i(t))^2}$. Table \ref{tab:baseline} \begin{table}[t]\centering
\caption{Comparison of Coordination Policies}
\label{tab:baseline}
\footnotesize\setlength{\tabcolsep}{4pt}
\begin{tabular}{lcccc}\toprule
Policy & RMSE (kW) & Cost (\$) & $\min_i|V_i|$ & Violations\\\midrule
Uncoordinated & $453.2$ & $869.6$ & $0.9353$ & $142$\\
No grid objective & $371.2$ & $874.1$ & $0.9500$ & $0$\\
Proposed & $19.7$ & $825.8$ & $0.9500$ & $0$\\
\bottomrule\end{tabular}
\\[2pt]{\footnotesize Cost is the negated prosumer payoff of \eqref{eq:payoffs};
violations are node-hours with $|V_i|<0.95$ p.u.}
\vspace*{-1mm}
\end{table}summarizes the outcome, and Figs. \ref{fig:tracking}  and \ref{fig:voltage} present the tracking and voltage results.

\begin{figure}[!htb]
\centering
\vspace*{-3mm}
\includegraphics[width=\columnwidth]{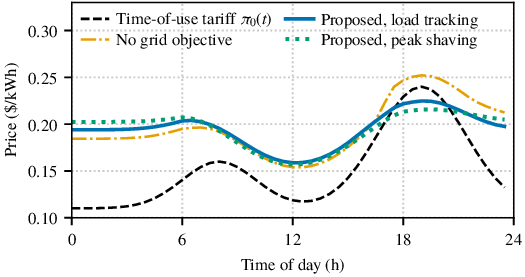}
\vspace*{-7mm}
\caption{Time-of-use tariff and node-averaged nodal price over the day under each grid-level objective.}
\vspace*{-2mm}
\label{fig:prices}
\end{figure}

The uncoordinated policy readily shows the impacts that distributed PV and EVs have on the feeder, i.e., concentrating charging in the overnight and late-evening hours when the time-of-use tariff $\pi_0(t)$, shown in Fig. \ref{fig:prices}, is low. 
 This drives the minimum voltage to $0.9353$ p.u. and takes $142$ node-hours below the $0.95$ p.u. limit, while missing the tracking target by $453$ kW RMSE. Enforcing the voltage constraints alone removes every violation but leaves the tracking error at $371$ kW. The proposed method reduces the tracking error to $19.7$ kW, a $95.7\%$ improvement over uncoordinated dispatch, while holding every node at or above the limit.

Notably, the proposed policy also has the lowest prosumer cost of the three, i.e., tracking is not obtained at the expense of the prosumers. The coordinating price moves charging into hours that are both cheaper and better for the feeder, so the grid-level objective and the private objectives are aligned rather than traded off at this penetration level.

\subsection{Other Grid-Level Objectives}
\label{subsec:otherobj}
The proposed pricing scheme is not limited to load tracking. To demonstrate its generality, we change only the function $g(\cdot)$ and solve \eqref{eq:P} again with no other modifications. As shown in Fig. \ref{fig:tracking}, peak shaving reduces the peak PCC import from $3638$ kW to $2104$ kW, a $42.2\%$ reduction. The same objective also fills the valley: The lowest PCC import rises from $549$ kW to $1989$ kW, so the daily swing shrinks from $3089$ kW to $114$ kW. Ramp limiting is visible in the same runs: the largest half-hourly change in PCC import falls from $463$ kW under uncoordinated dispatch to $244$ kW under load tracking and to $14$ kW under the peak-shaving objective.

Fig. \ref{fig:prices} shows how each objective reshapes the price. The node-averaged nodal price lies above the tariff because the energy price in \eqref{eq:energyprice} also contains the term $2c_0p_0^\star(t)$. Without a grid objective, the price follows the shape of the tariff and peaks at $0.252$ \$/kWh at around 19:00. Load tracking lowers the evening peak to $0.225$ \$/kWh and raises the overnight price by about $0.01$ \$/kWh, which moves charging into the evening. Peak shaving flattens the price further, and the daily price range shrinks from $0.098$ \$/kWh to $0.059$ \$/kWh. In every case, only the rule by which the aggregator computes $\gamma(t)$ changes, while the structure of \eqref{eq:price} and the local problems of the prosumers stay the same.

\subsection{EV Penetration and the Composition of the Price}
\label{subsec:penetration}
\begin{figure}[!htb]
\centering
\vspace*{-3mm}
\includegraphics[width=\columnwidth]{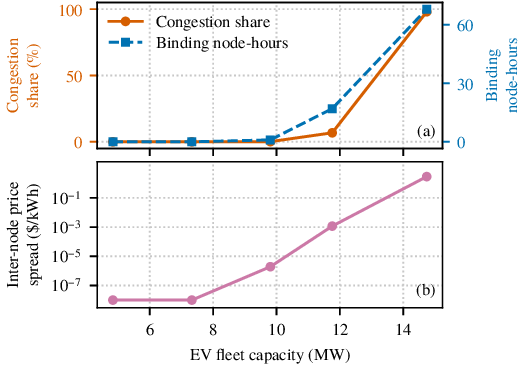}
\vspace*{-7mm}
\caption{Effect of EV penetration on the nodal price.} 
\vspace*{-2mm}
\label{fig:penetration}
\end{figure}

The dominant component of the nodal price \eqref{eq:price} depends on which resource is scarce. To examine the dependence, Fig. \ref{fig:penetration} sweeps the aggregated EV charging capacity of the feeder from $4.8$ MW to $14.7$ MW, which is about $0.5$ to $1.5$ times the base fleet of $9.81$ MW. As in Section \ref{subsec:case}, the fleet of each prosumer remains proportional to its nominal demand. Up to about $9.8$ MW, the voltage constraint is essentially inactive, the congestion component stays below $0.2\%$ of the peak price, and the prices are uniform across nodes to within $2\times10^{-6}$ \$/kWh, indicating that the feeder is not the binding resource and the price is almost entirely energy plus grid objective. Beyond the $9.8$ MW threshold, the voltage constraint starts to bind. At $11.8$ MW, the congestion component reaches $7\%$ of the peak price at $17$ binding node-hours. At $14.7$ MW, it reaches $98\%$ at $68$ binding node-hours, with an inter-node spread of $2.88$ \$/kWh.

\begin{figure}[!htb]
\centering
\vspace*{-3mm}
\includegraphics[width=\columnwidth]{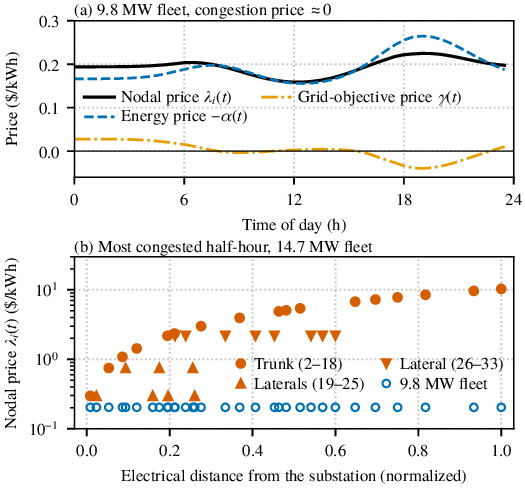}
\vspace*{-7mm}
\caption{Anatomy of the nodal price.}
\vspace*{-3mm}
\label{fig:price}
\end{figure}

Fig. \ref{fig:price} illustrates how the three components of the nodal price \eqref{eq:price} evolve over the day and vary across the feeder. Panel (a) shows how the grid-level objective enters the price, and panel (b) shows how the network constraints enter it. Panel (a) shows the base case of Section \ref{subsec:case}, i.e., the $9.8$ MW fleet. It can be readily observed that the congestion component is approximately zero, and the price is the common energy component plus the common scalar $\gamma(t)$. Around the evening peak of the energy price, $\gamma(t)$ becomes negative, lowering the nodal price and moving charging into the evening. This raises the PCC import toward the target in Fig. \ref{fig:tracking}. The scalar $\gamma(t)$ is the only part of the price that the grid-level objective controls. At the most congested half-hour, panel (b), the price is uniform across all $32$ buses at $9.8$ MW, whereas at $14.7$ MW it ranges from $0.29$ \$/kWh to $10.31$ \$/kWh and rises with electrical distance from the substation. The residual scatter at a given distance is not noise. By \eqref{eq:price}, the congestion term sums each binding node's sensitivity $R_{ki}$, so two buses equally far from the substation are priced differently when different limits are active behind them. By complementary slackness, the congestion component is nonzero only when at least one voltage limit binds. The same price formula therefore gives nearly uniform prices on an uncongested feeder and strongly differentiated prices once voltage headroom becomes scarce, and Fig. \ref{fig:penetration} shows that the transition between the two regimes is sharp. The comparison also shows why a formulation that omits the network, as several prior schemes \cite{khaki2018hierarchical,nimalsiri2021coordinated} do, is adequate at low EV penetration but misleading at high penetration.

\section{Conclusion}
\label{sec:conclusion}
This paper presented a distributed pricing scheme for multi-prosumer feeders that embeds the marginal value of a mixed competitive-cooperative grid-level objective into nodal prices. We proved that the centralized social optimum has zero duality gap and is exactly equivalent to a CE whose price decomposes into a system-wide energy component, a locational congestion component, and a grid-objective component. The grid-objective component is conveyed by a single broadcast scalar per time step. A projected AHU algorithm with an explicit step-size rule computes the CE at an $\mathcal{O}(1/\sqrt{L})$ ergodic rate, and its extragradient variant attains $\mathcal{O}(1/L)$. On the IEEE 33-bus feeder, the decentralized equilibrium reproduced the centralized optimum to $2\times10^{-6}$ kW, cut the PCC tracking error by $95.7\%$, and removed the $142$ node-hours of undervoltage incurred by uncoordinated dispatch, at the lowest prosumer cost of the three policies compared and with prices that decompose into interpretable components.


{\appendices
\section{Proof of Theorem 1}\label{proof_Th_1}
Because \eqref{eq:P} is convex and it satisfies Slater's condition, its KKT conditions are necessary and sufficient for global optimality, so $\bz^\star$ solves \eqref{eq:P} if and only if it satisfies that KKT system. Let $\balpha^\star$ and $\overline{\bxi}^\star,\underline{\bxi}^\star\ge0$ denote the multipliers of the coupling constraints \eqref{eq:bal} and \eqref{eq:volt}, respectively. Let $\bpsi_i^\star\ge0$ denote the multiplier of the surplus constraint in \eqref{eq:surplus}, and $\overline{\brho}_i^\star,\underline{\brho}_i^\star\ge0$ those of the SoC limits \eqref{eq:socbox}. The boxes \eqref{eq:ubox}, \eqref{eq:pbox}, and \eqref{eq:p0box} are handled through their normal cones $N(\cdot)$. The KKT conditions of \eqref{eq:P} are then, for all $i\in\N$ and $\tau\in\T$,
\begin{align}
0&\in f_i'(u_i^\star(\tau))+\Phi_i'(x_i^\star(T))\eta_i\Delta-\psi_i^\star(\tau)\notag\\
&\quad-\eta_i\Delta\!\sum_{t>\tau}\!\left(\overline{\rho}_i^\star(t)-\underline{\rho}_i^\star(t)\right)-N_{[-\overline{u}_i,\overline{u}_i]}(u_i^\star(\tau)),\label{eq:kkt_u}\\
0&\in \gamma^\star(\tau)-\alpha^\star(\tau)+\!\sum_{k\in\N}\!\left(\underline{\xi}_k^\star(\tau)-\overline{\xi}_k^\star(\tau)\right)R_{ki}-\psi_i^\star(\tau)\notag\\
&\quad-N_{[-\overline{p}_i,\overline{p}_i]}(p_i^\star(\tau)),\label{eq:kkt_p}\\
0&\in -C_0'(p_0^\star(\tau))-\alpha^\star(\tau)-N_{[\underline{p}_0,\overline{p}_0]}(p_0^\star(\tau)),\label{eq:kkt_p0}
\end{align}
together with primal feasibility, i.e., \eqref{eq:dyn}, \eqref{eq:devbox}, \eqref{eq:surplus}, \eqref{eq:volt}, and \eqref{eq:balance}, dual feasibility, i.e., $\bpsi^\star,\brho^\star,\bxi^\star\ge0$, and complementary slackness for \eqref{eq:surplus}, \eqref{eq:devbox}, and \eqref{eq:volt}. Here $\gamma^\star(\tau)=2\mu_g(P_b(\tau)-\sum_j p_j^\star(\tau))$ is the partial derivative of $g(\cdot)$, which is the same for every $i$, and we have used $\partial g/\partial p_i=\gamma^\star$ in \eqref{eq:kkt_p}.

By \eqref{eq:price}, the first three terms in \eqref{eq:kkt_p} sum to $\lambda_i^\star(\tau)$. Hence, \eqref{eq:kkt_p} can be rewritten as
\begin{equation}\label{eq:kkt_lam}
0\in\lambda_i^\star(\tau)-\psi_i^\star(\tau)-N_{[-\overline{p}_i,\overline{p}_i]}(p_i^\star(\tau)).
\end{equation}

\emph{Sufficiency.} Let $\bz^\star$ solve \eqref{eq:P}. Then $\bz^\star$ satisfies \eqref{eq:kkt_u}, \eqref{eq:kkt_p0} and \eqref{eq:kkt_lam} with the multipliers defined above. We verify conditions (i)--(iv) of Definition \ref{def:ce} in turn. For condition (i), fix a prosumer $i$ and consider \eqref{eq:LP} at the price $\blambda_i^\star$. The KKT conditions of \eqref{eq:LP} consist of \eqref{eq:kkt_u}, \eqref{eq:kkt_lam}, the local constraints \eqref{eq:dyn}, \eqref{eq:devbox} and \eqref{eq:surplus}, and complementary slackness for \eqref{eq:surplus} and \eqref{eq:socbox}. Each of them is also a KKT condition of \eqref{eq:P}, so all of them hold at $(\bu_i^\star,\bp_i^\star)$ with the multipliers $\bpsi_i^\star$, $\overline{\brho}_i^\star$ and $\underline{\brho}_i^\star$. The objective of \eqref{eq:LP} is concave by (A1) and its constraints are affine, so its KKT conditions are sufficient for optimality. Hence, $(\bu_i^\star,\bp_i^\star)$ solves \eqref{eq:LP}. Condition (ii) holds because \eqref{eq:kkt_p0} is the optimality condition of the concave substation problem in condition (ii). Condition (iii) holds by the definition of $\gamma^\star(t)$ in \eqref{eq:gamma}. Condition (iv) follows from primal feasibility of \eqref{eq:bal} and \eqref{eq:volt} and from complementary slackness for \eqref{eq:volt}. Therefore, $\bz^\star$ and the prices \eqref{eq:price} form a CE.

\emph{Necessity.} Let $(\{\blambda_i^\star\},\balpha^\star,\Xi^\star,\bgamma^\star,\bz^\star)$ be a CE. By condition (i) of Definition \ref{def:ce}, $(\bu_i^\star,\bp_i^\star)$ solves \eqref{eq:LP} for each $i$. The constraints of \eqref{eq:LP} are affine, so its KKT conditions are necessary for optimality. Hence, there exist $\bpsi_i^\star\ge0$ and $\overline{\brho}_i^\star,\underline{\brho}_i^\star\ge0$ such that \eqref{eq:kkt_u} and \eqref{eq:kkt_lam} hold, together with complementary slackness for \eqref{eq:surplus} and \eqref{eq:socbox}. Every $\blambda_i^\star$ is built from the same $\balpha^\star$, $\Xi^\star$, and $\bgamma^\star$, and, by condition (iii), $\gamma^\star(\tau)$ equals the partial derivative \eqref{eq:gderiv} of $g$ evaluated at $\bp^\star(\tau)$. Substituting \eqref{eq:price} into \eqref{eq:kkt_lam} therefore gives \eqref{eq:kkt_p} with multipliers that are common to all prosumers. Condition (ii) gives \eqref{eq:kkt_p0}, and condition (iv) gives primal feasibility of \eqref{eq:bal} and \eqref{eq:volt}, dual feasibility $\Xi^\star\ge0$, and complementary slackness for \eqref{eq:volt}. All KKT conditions of \eqref{eq:P} therefore hold at $\bz^\star$. Since the KKT conditions of \eqref{eq:P} are sufficient, $\bz^\star$ solves \eqref{eq:P}.


\section{Proof of Theorem 2}\label{proof_Th_2}
The Lagrangian used by Algorithm 1 can be written as $\mathcal{L}(\bz,\bmu)=W(\bz)-\bmu^{\mathsf{T}}\bc(\bz)$, where $\bc(\bz)$ stacks the residuals of the dualized constraints \eqref{eq:bal}, \eqref{eq:volt}, \eqref{eq:surplus} and \eqref{eq:socbox}, written so that the inequality residuals are nonpositive at a feasible point. Hence, $\mathcal{L}(\bz,\bmu)$ is concave in $\bz$ and affine in $\bmu$. Let $\bz\in\mathcal{Z}$ and $\bmu\in\mathcal{M}_\rho$ be arbitrary. First, we show that, for every $\ell$,
\begin{equation}\label{eq:pstep}
\mathcal{L}(\bz,\bmu^\ell){-}\mathcal{L}(\bz^\ell,\bmu^\ell){\le}\frac{\|\bz^\ell{-}\bz\|_2^2{-}\|\bz^{\ell+1}{-}\bz\|_2^2}{2\beta_p}+\frac{\beta_pG^2}{2}.
\end{equation}
Let $\bg^\ell{=}\nabla_{\bz}\mathcal{L}(\bz^\ell,\bmu^\ell)$, so that $\bz^{\ell+1}{=}\Pi_{\mathcal{Z}}(\bz^\ell+\beta_p\bg^\ell)$. Since $\bz\in\mathcal{Z}$ and the projection onto $\mathcal{Z}$ is nonexpansive, we have
\begin{equation}\label{eq:zstep}
\|\bz^{\ell+1}-\bz\|_2^2\le\|\bz^\ell-\bz\|_2^2+2\beta_p(\bz^\ell-\bz)^{\mathsf{T}}\bg^\ell+\beta_p^2\|\bg^\ell\|_2^2 .
\end{equation}
Since $\mathcal{L}(\cdot,\bmu^\ell)$ is concave, we have
\begin{equation}\label{eq:concave}
(\bz^\ell-\bz)^{\mathsf{T}}\bg^\ell\le\mathcal{L}(\bz^\ell,\bmu^\ell)-\mathcal{L}(\bz,\bmu^\ell).
\end{equation}
Substituting \eqref{eq:concave} and $\|\bg^\ell\|_2\le G$ from Lemma \ref{lem:bound} into \eqref{eq:zstep} and dividing both sides by $2\beta_p$ give \eqref{eq:pstep}.

 We then show that, for every $\ell$,
\begin{align}\label{eq:dstep}
&\mathcal{L}(\bz^{\ell+1},\bmu^\ell)-\mathcal{L}(\bz^{\ell+1},\bmu)\notag\\
&\qquad\le\frac{\|\bmu^\ell-\bmu\|_2^2-\|\bmu^{\ell+1}-\bmu\|_2^2}{2\beta_d}+\frac{\beta_dG^2}{2}.
\end{align}
Let $\bh^\ell{=}\nabla_{\bmu}\mathcal{L}(\bz^{\ell+1},\bmu^\ell)$, so that $\bmu^{\ell+1}=\Pi_{\mathcal{M}}(\bmu^\ell-\beta_d\bh^\ell)$. Since $\bmu\in\mathcal{M}$ and $\Pi_{\mathcal{M}}$ is nonexpansive, we have
\begin{equation}\label{eq:mustep}
\|\bmu^{\ell+1}-\bmu\|_2^2\le\|\bmu^\ell-\bmu\|_2^2-2\beta_d(\bmu^\ell-\bmu)^{\mathsf{T}}\bh^\ell+\beta_d^2\|\bh^\ell\|_2^2 .
\end{equation}
Since $\mathcal{L}(\bz^{\ell+1},\cdot)$ is affine, there holds
\begin{equation}\label{eq:affine}
(\bmu^\ell-\bmu)^{\mathsf{T}}\bh^\ell=\mathcal{L}(\bz^{\ell+1},\bmu^\ell)-\mathcal{L}(\bz^{\ell+1},\bmu).
\end{equation}
Substituting \eqref{eq:affine} and $\|\bh^\ell\|_2\le G$ into \eqref{eq:mustep}, rearranging the terms, and dividing both sides by $2\beta_d$ give \eqref{eq:dstep}.

The bounds \eqref{eq:pstep} and \eqref{eq:dstep} are evaluated at $\bz^\ell$ and $\bz^{\ell+1}$, respectively. To combine them, we bound the difference between these two points as
\begin{equation}\label{eq:cross}
\mathcal{L}(\bz^{\ell},\bmu^\ell)-\mathcal{L}(\bz^{\ell+1},\bmu^\ell)\le G\|\bz^{\ell+1}-\bz^\ell\|_2\le\beta_pG^2 .
\end{equation}
The first inequality holds because $\|\bg^\ell\|_2\le G$ makes $\mathcal{L}(\cdot,\bmu^\ell)$ Lipschitz with constant $G$ on $\mathcal{Z}$. The second inequality holds because the projection is nonexpansive, so $\|\bz^{\ell+1}{-}\bz^\ell\|_2{\le}\beta_p\|\bg^\ell\|_2{\le}\beta_pG$. The left-hand sides of \eqref{eq:pstep}, \eqref{eq:cross}, and \eqref{eq:dstep} add up to $\mathcal{L}(\bz,\bmu^\ell)-\mathcal{L}(\bz^{\ell+1},\bmu)$. Adding the three inequalities, therefore, gives
\begin{align}\label{eq:onestep}
&\mathcal{L}(\bz,\bmu^\ell)-\mathcal{L}(\bz^{\ell+1},\bmu)\le\frac{\|\bz^\ell-\bz\|_2^2-\|\bz^{\ell+1}-\bz\|_2^2}{2\beta_p}\notag\\
&\quad+\frac{\|\bmu^\ell-\bmu\|_2^2-\|\bmu^{\ell+1}-\bmu\|_2^2}{2\beta_d}+\frac{(3\beta_p+\beta_d)G^2}{2}.
\end{align}

 Summing \eqref{eq:onestep} over $\ell=0,\dots,L-1$ cancels $\|\bz^{\ell}-\bz\|_2^2$ for $1\le\ell\le L-1$, leaving only $\|\bz^0-\bz\|_2^2-\|\bz^L-\bz\|_2^2$. The same holds for $\bmu$. We drop the nonpositive terms $-\|\bz^L-\bz\|_2^2$ and $-\|\bmu^L-\bmu\|_2^2$ and divide both sides of the sum by $L$. Since $\mathcal{L}(\bz,\cdot)$ is affine, the average of $\mathcal{L}(\bz,\bmu^\ell)$ over $\ell$ equals $\mathcal{L}(\bz,\bar\bmu^{L})$. Since $\mathcal{L}(\cdot,\bmu)$ is concave, Jensen's inequality shows that the average of $\mathcal{L}(\bz^{\ell+1},\bmu)$ over $\ell$ is at most $\mathcal{L}(\bar\bz^{L},\bmu)$. Together, these give
\begin{align}\label{eq:gapraw}
&\mathcal{L}(\bz,\bar\bmu^{L})-\mathcal{L}(\bar\bz^{L},\bmu)\notag\\
&\quad\le\frac{\|\bz^0-\bz\|_2^2}{2\beta_pL}+\frac{\|\bmu^0-\bmu\|_2^2}{2\beta_dL}+\frac{(3\beta_p+\beta_d)G^2}{2}.
\end{align}

 With $\beta_p=\beta_d=\beta$, the first two terms in \eqref{eq:gapraw} are at most $D^2/(2\beta L)$ by \eqref{eq:D}, and the last term equals $2\beta G^2$. Hence, the right-hand side of \eqref{eq:gapraw} is at most
\begin{equation}\label{eq:betafun}
\phi(\beta)=\frac{D^2}{2\beta L}+2\beta G^2 .
\end{equation}
The derivative $\phi'(\beta)=-D^2/(2\beta^2L)+2G^2$ vanishes when $\beta$ satisfies \eqref{eq:step}, which therefore minimizes $\phi(\cdot)$, yielding
\begin{equation}\label{eq:betaval}
\phi(\beta)|_{{D}/{(2G\sqrt{L})}}={2DG}/{\sqrt{L}}.
\end{equation}
Taking the supremum of the left-hand side of \eqref{eq:gapraw} over $\bz\in\mathcal{Z}$ and $\bmu\in\mathcal{M}_\rho$ gives \eqref{eq:gap}.

Let $\epsilon_L=2DG/\sqrt{L}$, and let $\bc^+(\bz)$ keep the equality residuals of the residual vector $\bc(\bz)$ defined at the start of the proof and replace each inequality residual by its positive part, so that $\|\bc^+(\bz)\|_2$ measures the constraint violation of $\bz$. Since $\bz^\star$ is feasible and $\bar\bmu^{L}\in\mathcal{M}$, we have $\mathcal{L}(\bz^\star,\bar\bmu^{L})\ge W^\star$. Setting $\bz=\bz^\star$ in \eqref{eq:gap} therefore gives
\begin{equation}\label{eq:feas}
W^\star-W(\bar\bz^{L})+\bmu^{\mathsf{T}}\bc(\bar\bz^{L})\le\epsilon_L,\qquad\forall\bmu\in\mathcal{M}_\rho .
\end{equation}
Since $(\bz^\star,\bmu^\star)$ is a saddle point, $\mathcal{L}(\bar\bz^{L},\bmu^\star)\le\mathcal{L}(\bz^\star,\bmu^\star)=W^\star$, which gives
\begin{equation}\label{eq:lowerW}
W^\star-W(\bar\bz^{L})+(\bmu^\star)^{\mathsf{T}}\bc(\bar\bz^{L})\ge0 .
\end{equation}
To bound the infeasibility, we choose $\bmu=\bmu^\star+\frac{\rho}{2}\bc^+(\bar\bz^{L})/\|\bc^+(\bar\bz^{L})\|_2$ in \eqref{eq:feas}. This $\bmu$ lies in $\mathcal{M}_\rho$ because its inequality components are nonnegative and $\|\bmu\|_2\le\|\bmu^\star\|_2+\rho/2\le\rho$. Since $\bc^+(\bar\bz^{L})^{\mathsf{T}}\bc(\bar\bz^{L})=\|\bc^+(\bar\bz^{L})\|_2^2$, combining \eqref{eq:feas} for this $\bmu$ with \eqref{eq:lowerW} gives
\begin{equation}\label{eq:infeas}
\|\bc^+(\bar\bz^{L})\|_2\le\frac{2\epsilon_L}{\rho}=\frac{4DG}{\rho\sqrt{L}} .
\end{equation}
To bound the suboptimality, setting $\bmu=0$ in \eqref{eq:feas} gives $W^\star-W(\bar\bz^{L})\le\epsilon_L$. Moreover, \eqref{eq:lowerW}, the Cauchy-Schwarz inequality, \eqref{eq:infeas}, and $\rho\ge2\|\bmu^\star\|_2$ give $W^\star-W(\bar\bz^{L})\ge-\|\bmu^\star\|_2\|\bc^+(\bar\bz^{L})\|_2\ge-\epsilon_L$. Hence, $|W^\star-W(\bar\bz^{L})|\le\epsilon_L$. Finally, the averages lie in the compact set $\mathcal{Z}\times\mathcal{M}_\rho$, so convergent subsequences exist. Let $(\hat\bz,\hat\bmu)$ be the limit of one of them. By \eqref{eq:infeas} and the suboptimality bound, $\hat\bz$ is feasible and $W(\hat\bz)=W^\star$, so $\hat\bz$ solves \eqref{eq:P}. Passing to the limit in \eqref{eq:gap} with $\bmu=0$ gives $\max_{\bz\in\mathcal{Z}}\mathcal{L}(\bz,\hat\bmu)\le W(\hat\bz)=W^\star$, which by weak duality indicates that $\hat\bmu$ is an optimal multiplier.

\section{Proof of Theorem 3}\label{Proof_Th_3}
The argument follows the standard extragradient analysis \cite{nemirovski2004prox}. Let $\bw={[\bz^{\mathsf{T}},\bmu^{\mathsf{T}}]}^{\mathsf{T}}$, $\hat\bw^\ell={[(\hat\bz^\ell)^{\mathsf{T}},(\hat\bmu^\ell)^{\mathsf{T}}]}^{\mathsf{T}}$ and $\mathcal{W}=\mathcal{Z}\times\mathcal{M}$. Since the projections onto $\mathcal{Z}$ and $\mathcal{M}$ act on separate blocks, \eqref{eq:eg} can be written as
\begin{equation}\label{eq:egw}
\hat\bw^\ell{=}\Pi_{\mathcal{W}}\bigl(\bw^\ell{-}\beta F(\bw^\ell)\bigr), \bw^{\ell+1}{=}\Pi_{\mathcal{W}}\bigl(\bw^\ell{-}\beta F(\hat\bw^\ell)\bigr).
\end{equation}
 Since $\mathcal{L}$ is concave in $\bz$ and affine in $\bmu$, $F$ is monotone, i.e., $\bigl(F(\bw)-F(\bw')\bigr)^{\mathsf{T}}(\bw-\bw')\ge0$ for all $\bw,\bw'\in\mathcal{W}$. Moreover, as in the proof of Theorem \ref{thm:conv}, $\mathcal{L}(\bz,\bmu)=W(\bz)-\bmu^{\mathsf{T}}\bc(\bz)$, where $W$ is quadratic and $\bc(\bz)=\mathcal{A}\bz+\bc_0$ is affine with a constant vector $\bc_0$. Hence, for all $\bw,\bw'\in\mathcal{W}$, we have
\begin{equation}\label{eq:Faffine}
F(\bw)-F(\bw')=\begin{bmatrix}-\nabla^2_{\bz\bz}\mathcal{L} & \mathcal{A}^{\mathsf{T}}\\ -\mathcal{A} & 0\end{bmatrix}(\bw-\bw').
\end{equation}
The matrix in \eqref{eq:Faffine} is the sum of a block-diagonal part with norm $\|\nabla^2_{\bz\bz}\mathcal{L}\|_2$ and a skew-symmetric part with norm $\|\mathcal{A}\|_2$. Therefore, $F$ is Lipschitz on all of $\mathcal{W}$ with constant $\Lambda$.

 We show that, for every $\bw\in\mathcal{W}$ and every $\ell$,
\begin{equation}\label{eq:egstep}
\beta F(\hat\bw^\ell)^{\mathsf{T}}(\hat\bw^\ell-\bw)\le\frac{\|\bw^\ell-\bw\|_2^2-\|\bw^{\ell+1}-\bw\|_2^2}{2}.
\end{equation}
For a closed convex set $\mathcal{S}$, the projection $\boldsymbol{y}=\Pi_{\mathcal{S}}(\boldsymbol{x})$ satisfies $(\boldsymbol{x}-\boldsymbol{y})^{\mathsf{T}}(\boldsymbol{v}-\boldsymbol{y})\le0$ for all $\boldsymbol{v}\in\mathcal{S}$. Applying this property to the corrector in \eqref{eq:egw} with $\boldsymbol{v}=\bw$ and to the predictor with $\boldsymbol{v}=\bw^{\ell+1}$ gives
\begin{align}
\beta F(\hat\bw^\ell)^{\mathsf{T}}(\bw^{\ell+1}-\bw)\le&\frac{1}{2}\left(\|\bw^\ell-\bw\|_2^2-\|\bw^{\ell+1}-\bw\|_2^2 \right.\notag\\
&\quad\left. -\|\bw^\ell-\bw^{\ell+1}\|_2^2\right),\label{eq:egA}\\
\beta F(\bw^\ell)^{\mathsf{T}}(\hat\bw^\ell-\bw^{\ell+1})\le&\frac{1}{2}\left(\|\bw^\ell-\bw^{\ell+1}\|_2^2-\|\bw^\ell -\hat\bw^\ell\|_2^2 \right.\notag\\
&\quad \left.-\|\hat\bw^\ell-\bw^{\ell+1}\|_2^2\right).\label{eq:egB}
\end{align}
By the Cauchy-Schwarz inequality and the Lipschitz property of $F(\cdot)$ established above, we have
\begin{align}\label{eq:egC}
&\beta\bigl(F(\hat\bw^\ell)-F(\bw^\ell)\bigr)^{\mathsf{T}}(\hat\bw^\ell-\bw^{\ell+1})\notag\\
&\quad\le\frac{1}{2}\beta^2\Lambda^2\|\hat\bw^\ell-\bw^\ell\|_2^2+\frac{1}{2}\|\hat\bw^\ell-\bw^{\ell+1}\|_2^2 .
\end{align}
The left-hand sides of \eqref{eq:egA}, \eqref{eq:egB}, and \eqref{eq:egC} add up to the left-hand side of \eqref{eq:egstep}. Adding the three inequalities, the terms $\|\bw^\ell-\bw^{\ell+1}\|_2^2$ and $\|\hat\bw^\ell-\bw^{\ell+1}\|_2^2$ cancel, yielding
\begin{align}\label{eq:egsum}
\beta F(\hat\bw^\ell)^{\mathsf{T}}(\hat\bw^\ell-\bw)\le&\frac{\|\bw^\ell-\bw\|_2^2-\|\bw^{\ell+1}-\bw\|_2^2}{2}\notag\\
&\quad-\frac{1-\beta^2\Lambda^2}{2}\|\hat\bw^\ell-\bw^\ell\|_2^2 .
\end{align}
Since $\beta\le1/(\sqrt{2}\,\Lambda)$, we have $1-\beta^2\Lambda^2>0$, so dropping the last term of \eqref{eq:egsum} gives \eqref{eq:egstep}.

We then show that, for every $\bw\in\mathcal{W}$,
\begin{equation}\label{eq:eglink}
F(\hat\bw^\ell)^{\mathsf{T}}(\hat\bw^\ell-\bw)\ge\mathcal{L}(\bz,\hat\bmu^\ell)-\mathcal{L}(\hat\bz^\ell,\bmu).
\end{equation}
By the definition of $F(\cdot)$, the left-hand side of \eqref{eq:eglink} equals $\nabla_{\bz}\mathcal{L}(\hat\bz^\ell,\hat\bmu^\ell)^{\mathsf{T}}(\bz-\hat\bz^\ell)+\nabla_{\bmu}\mathcal{L}(\hat\bz^\ell,\hat\bmu^\ell)^{\mathsf{T}}(\hat\bmu^\ell-\bmu)$. Since $\mathcal{L}(\cdot,\hat\bmu^\ell)$ is concave, the first term is at least $\mathcal{L}(\bz,\hat\bmu^\ell)-\mathcal{L}(\hat\bz^\ell,\hat\bmu^\ell)$. Since $\mathcal{L}(\hat\bz^\ell,\cdot)$ is affine, the second term equals $\mathcal{L}(\hat\bz^\ell,\hat\bmu^\ell)-\mathcal{L}(\hat\bz^\ell,\bmu)$. Adding the lower bound on the first term and the expression for the second term gives \eqref{eq:eglink}.

 Let $\bz\in\mathcal{Z}$ and $\bmu\in\mathcal{M}_\rho$ be arbitrary. Combining \eqref{eq:eglink} with \eqref{eq:egstep} and summing over $\ell=0,\dots,L-1$, the distance terms cancel in pairs except $\|\bw^0-\bw\|_2^2-\|\bw^L-\bw\|_2^2$. We drop the nonpositive term $-\|\bw^L-\bw\|_2^2$ and divide both sides by $\beta L$. Since $\mathcal{L}(\bz,\cdot)$ is affine, the average of $\mathcal{L}(\bz,\hat\bmu^\ell)$ over $\ell$ equals $\mathcal{L}(\bz,\bar\bmu^{L})$. Since $\mathcal{L}(\cdot,\bmu)$ is concave, Jensen's inequality shows that the average of $\mathcal{L}(\hat\bz^\ell,\bmu)$ over $\ell$ is at most $\mathcal{L}(\bar\bz^{L},\bmu)$. Moreover, $\|\bw^0-\bw\|_2^2=\|\bz^0-\bz\|_2^2+\|\bmu^0-\bmu\|_2^2\le D^2$ by \eqref{eq:D}. Together, these give
\begin{equation}\label{eq:egfinal}
\mathcal{L}(\bz,\bar\bmu^{L})-\mathcal{L}(\bar\bz^{L},\bmu)\le\frac{D^2}{2\beta L}.
\end{equation}
Taking the supremum of the left-hand side of \eqref{eq:egfinal} over $\bz\in\mathcal{Z}$ and $\bmu\in\mathcal{M}_\rho$ gives \eqref{eq:eggap}. Unlike Theorem \ref{thm:conv}, this argument does not require the dual iterates to stay bounded, because $F(\cdot)$ is Lipschitz on $\mathcal{W}$.
}


\bibliographystyle{IEEEtran}
\bibliography{refs}

\end{document}